# A Brief Survey And Investigation Of Hybrid Beamforming For Millimeter Waves In 5G Massive MIMO Systems

Qazwan Abdullah[1,*], Adeb Salh[1], Nor Shahida Mohd Shah[2,*], Noorsaliza Abdullah[1], Lukman Audah[1], Shipun Anuar Hamzah[1], Nabil Farah[3], Maged Aboali[3], and Shahilah Nordin[4]

[1]Faculty of Electrical and Electronic Engineering, Universiti Tun Hussein Onn Malaysia, Batu Pahat, Johor, Malaysia

[2]Faculty of Engineering Technology, Universiti Tun Hussein Onn Malaysia, Pagoh, Johor, Malaysia

[3]Faculty of Electrical Engineering, Universiti Teknikal Malaysia Melaka, Melaka, Malaysia

[4]Faculty of Electrical Engineering, Universiti Teknologi Mara, Permatang Pauh, Penang, Malaysia

E-mails (*): gazwan20062015@gmail.com, shahida@uthm.edu.my

*Abstract*— Millimeter-wave (mm-wave) is a promising technique to enhance the network capacity and coverage of next-generation (5G) based on utilizing a great number of available spectrum resources in mobile communication. Improving the 5G network requires enhancing and employing mm-wave beamforming channel propagation characteristics. To achieve high data rates, system performance remains a challenge given the impact of propagation channels in mm-wave that is insufficient in both path loss, delay spread, and penetration loss. Additional challenges arise due to high cost and energy consumption, which require combining both analog and digital beamforming (hybrid beamforming) to reduce the number of radio frequency (RF) chains. In this paper, the distributed powers in the small cell to suppress path loss by specifying a considerable power and controlling the distributed power to reduce the high cost and energy consumption was proposed. The hybrid beamforming in mm-wave exploits a large bandwidth which reduces the large path loss in Rayleigh fading channel. Also, the trade-off between the energy consumption of RF chains and cost efficiency depends on reducing the number of RF chains and the distributed number of users. This paper finds that hybrid beamforming for massive multiple-input multiple-output (MIMO) systems constitute a promising platform for advancing and capitalizing on 5G networks.

## 1. Introduction

Massive multiple-input-multiple-output (MIMO) antennas are moving towards the millimeter-wave (mm-wave) to increase high data rate and enhance spectral and energy efficiencies based on larger bandwidth in the mm-wave frequency bands. The high multiplexing gains are achievable with a huge number of antennas for both transmitter and receiver [1]. Increasing the large data to active users, obtaining better costs, and conventional energy efficiency are considered key components of 5G wireless communication systems. The improved spectral efficiency depends on using massive MIMO techniques at transmitting the signal from the base station (BS) to active users (UEs) simultaneously at the same frequency. Also, increasing the number of antennas allows increasing the data stream to a high value in every cell and simplifies signal processing due to small-scale fading. For 5G to improve wireless communication 1000-fold by 2020 for intelligent wireless facilities such as high video streaming, cell transmitting, and user equipment and, user equipment data traffic, massive MIMO systems can support this goal by reducing the noise and interference at the edge cell and reducing the power consumption of a digital to analog converter (DACs) that grows due to employing large antenna arrays. The sample of one-bit ADCs that consume little mill watts needs to compute the linear amplifier and control gain corresponding to radiofrequency (RF) chains [2-5].

The performance of the digital array and cost efficiency of analog arrays can be improved using the hybrid analog-digital by making the number of RF chains smaller than the number of antennas to decrease the required transmission energy. Employing large antenna arrays with the high cost and high consumption power of the mixed hybrid signal components are among the challenges associated with providing an individual RF chain for every antenna. Combining large analog with lower digital processing is called hybrid multiple antenna transceivers and can reduce hardware cost. The efficient number of RF chains can be obtained based on analyzing the diversity order for the number of antenna elements for hybrid beamforming [6], [7].

## 2. Hybrid Beamforming Channel

Large channel bandwidth and good communication links are evaluated by employing mm-wave beamforming to reduce the high path loss. The large mm-wave frequency band spans from 10GHz to 300 GHz that satisfies high data traffic. While the small wavelength for mm-wave can decrease the attenuation and increases the capacity by using the directional gain beneficial to great numbers of half-wavelengths which spread out antennas into a small area. The high achievable capacity and the improvement of bit error rate require using the diversity gain and multiplexing gain while mitigating co-channel interference and improving a signal to interference noise ratio requires antenna gain. The suppressed path loss depends on using achievable line-of-sight in a small cell and by applying less multi-path signals for uplink and downlink in a multi-tier heterogeneous system [3], [8-9]. The high directional beamforming for a large number of the antenna array in mm-wave massive MIMO is used to reduce the path loss with the directional transmission. The estimating channel based on using full channel information with hybrid beamforming can increase the high data rate for users. The hybrid beamforming in mm-wave exploits a large bandwidth to reduce the large path loss in Rayleigh fading channel. The uniform antenna array radiation patterns can suppress the interference power for SINR and increase the achievable data rates when the channel response vectors between UEs are orthogonal. Also, mitigating inter-beam interference depends on selecting conventional low RF chains and using digital precoding that decreases the co-channel interference from different UEs.

The full digital precoding in mm-wave beamforming makes it difficult to obtain a good channel because the full channel information used in the mm-wave is costly and incurs high-power consumption [10]. Hybrid beamforming is challenging as it entails coupling analog and digital beamforming to provide conventional low RF chains and low-cost of power consumption. The reciprocity in [11] is used to estimate channel state information in downlink based on assuming a model for propagation channel in hybrid beamforming massive MIMO. The accurate channel can be obtained in [11] based on used reverse channel estimation. The hybrid beamforming in [12] proposed maximizing the capacity for the baseband based on obtaining the near-optimal linear hybrid beamforming to a generic channel.

In this paper, low-cost power consumption was investigated by providing conventional low RF chains by distributed powers in the small cells to suppress path loss based on state a considerable power and controlling the distributed power with low complexity for Millimeter Wave.

### 3. Millimeter-Wave Propagation

Mm-wave channel propagation is critical technique for obtaining an accurate channel model to develop the 5G network for wideband wireless cellular networks. The characteristics of the channel in mm-wave is not sufficient due to high path loss, delay spread and penetration loss. The high directional gain antenna is used to decrease high propagation loss, where the non-line-of-sight needs a large high-frequency to decrease penetration loss. The square of the carrier frequency is proportional with line-of-sight (LoS) in free space path loss. The allowed frequency band in mm-wave ranges from 20 GHz to 300 GHz, which requires a high directional gain to decrease propagation loss. However, the used path loss at combination mm-wave with the device to the device causes the interference between users to be relatively low due to the employment of directional antennas. The characteristics of a combination mm-wave technique in a device to the device provide high capacity in a cellular network and large security anti eavesdropping [13]. The high penetration loss occurs when the user inside the house is connected with the outdoor base station at the transmit signal through the building wall, which discredits the data rate and energy consumption. The high-frequency band is planned to support 10 Gb/s as 60 GHz to achieve a high data rate and short-range. The band range from 57 GHz to 66 GHz is used in Europe, while the band range in the United States is from 57-64 GHz, and the band range available in Japan is from 59-63 GHz [14]. Small-scale fading is created due to the multipath propagation, and the path loss depends on signal power loss, which normally follows the power in the propagation medium. Authors in [15] proposed a new channel model to support 5G network for bands up to 100 GHz based on accessible three-dimensional of the 3[rd] Generation Partnership Project 3GPP 3D channel models. While the authors in [16] proposed two different frequency bands for path loss from 2 to 28 GHz and 2 to 73.5 GHz in an urban-macro environment to predict path loss at different distances. When transmitting a signal through the channel, the path loss decreases the power density which can be expressed as [17].

$$Q_{dB} = 10 \ \log_{10} \frac{Q_T}{Q_R} \quad (1)$$

where $Q_T$ is the transmit power, and $Q_R$ is the received power. The power in propagation medium for distance and wavelength can be written as

$$Q_R(d, \lambda) = \acute{G}_T \ \acute{G}_R \left(\frac{\lambda}{4\pi d}\right)^2 \quad (2)$$

where $\acute{G}_T$ is antenna gain for the transmitter, $\acute{G}_R$ is the antenna gain for the receiver, $\lambda$ is the wavelength and $d$ is the distance between transmitter and receiver. The very short wavelength allows the use of a large number of antenna arrays to achieve adequate beamforming gains, and the path loss is directly proportional to the propagation distance.

Research in this area can be categorized into two optimal solutions. Firstly, the investigation of mm-wave channel model to achieve high spectral efficiency. The authors in [18] compared mm-wave massive MIMO systems with several distributed antenna arrays by performing mm-wave based on distributed sub-array architecture. The author research in [18] focused on asymptotical maximum multiplexing gain analysis to achieve high performance. Table 1 shows multiple researches achieved high spectral efficiency based on the estimated channel.

The author in [19] studied the uniform linear array (ULA) and three sub-ULA configurations in base station (BS) to achieve spectral efficiency using the squared inner product between two-channel vectors to decrease the inter-user interference. While authors in [20] studied the spectral efficiency by using the learning, which allowed mm-wave massive MIMO for effective hybrid precoding to achieve high spectral efficiency for good performance in hybrid precoding. In [21], the authors focused on a simplified mm-wave channel by using concentration capacity analysis for the combined non-orthogonal multiple access (NOMA) technique with mm-wave-massive MIMO systems. Based on the capacity analysis, the author in [21] improved the capacity to achieve high spectral efficiency.

## 4. Low-Cost Hybrid Digital-Analogue Beamforming for Multiuser Massive MIMO

The Hybrid beamforming techniques exploit the full potential of a massive MIMO system by applying digital beamforming precoding techniques. The proposed hybrid beamforming improves the channel through the number of RF chains by reducing the hardware complexity by transmitting a large number of antennas and analog to digital converter.

TABLE 1. SUMMARY FOR ACHIEVING HIGH SPECTRAL EFFICIENCY BASED ON THE ESTIMATED CHANNEL

| Authors | Features | Limitation |
|---|---|---|
| Dain, W. et al | • Achievable spectral efficiency<br>• Employing distributed antenna sub-array architecture when the number of antennas goes to infinity | • Used multiplexing gain |
| Weijie, T. et al | • Achievable spectral efficiency based on mm-wave<br>• Estimate channel model based on a finite-dimensional mm-Wave channel by taking the azimuth angle | • Full uniform linear array<br>• Multiple sub-arrays antenna |
| Hongi, H. et al | • Deep learning-enabled mm-wave massive<br>• Minimizing the bit error ratio to enhance spectral efficiency | • Deep neural network (DNN) |
| Di, Z. et al | • Performing capacity analysis for the integrated NOMA- based on extending the uniform random single-path. | • Low SINR<br>• High SINR |

| | • Analyzing the capacity based on dominant factors of signal to interference plus noise ratio | • Probability density function (PDF) |
|---|---|---|

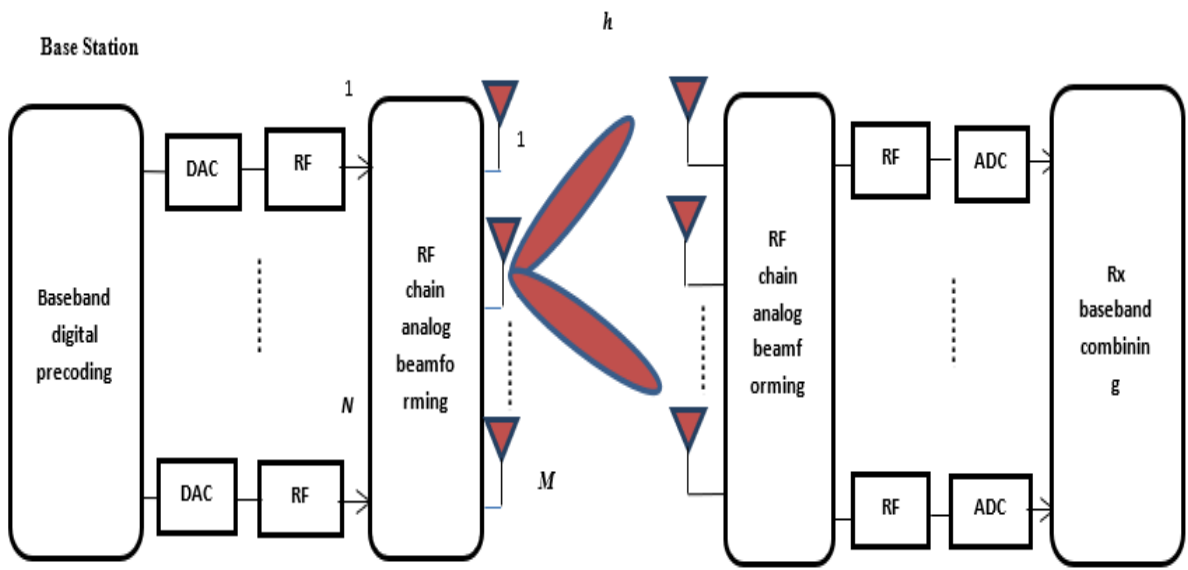

Figure1. Hybrid analogue-digital architectures in a massive MIMO channel.

Hybrid beamforming technique can be smaller by decreasing the number of multiplexed symbols. Hybrid beamforming technique consists of two parts. The first part is digital precoding or decoders, while the second part is analog precoders or decoders. Conjointly improving the digital precoders and combiner by exploiting a lower bound can achieve an average sum-rate based on using hybrid analog-digital beamforming [22]. In [22] the mm-wave adopted a hybrid RF chain by proposing hybrid beamforming that uses a very fewer number of RF chains to reduce the consumption circuit power and enact conventional digital beamforming. The mm-wave technique typically utilizes a small number of RF chains than a large number of transmit antennas with a massive MIMO system. To improve the designing hybrid analog/digital beamforming requires transmitting baseband precoding to be imposed by using fewer RF chains, while at the received signal in the uplink, RF chains can be followed by baseband decoding to generate low-cost hybrid beamforming. Moreover, analog RF chain beamforming can be implemented to achieve a high data rate by using phase shift, and digital baseband beamforming could be employed using microprocessing [23]. The high cost and power consumption of combined signal components could be reduced by preventing devoting an unconnected RF chain for every antenna based on employing a massive MIMO baseband precoding, as shown in Fig. 1. Both baseband and RF chains are critical designs for RF beamforming. The joint optimization RF chains can be obtained using fewer RF chains than a large number of transmit antennas. While the baseband beamforming is optimized by reducing the minimum mean square error MMSE among the information path and its evaluation under power constraints According to the combiner hardware in [24] which used a fully connected phase shifter, every antenna for the analog signal was connected to feed RF chains using on/off switch as a phase shifter.

Low-cost hybrid digital-analog beamforming is critical to reducing hardware complexity, where the connection depends on every RF chain being connected with all antennas at transmitting a signal for all antennas digital transceivers to decrease circuit power consumption due to the power amplifier, data converter, mixer, and phase shifter. The comprehensive connection is more practical to provide high complexity beamforming gain by enhancing the number of antennas per transceiver of the RF chain to improve energy efficiency and spectral efficiency. In [27], the trade-off between the energy consumption of RF chains and cost efficiency depends on reducing the number of RF chains and a distributed number of users. While sub-connection for hybrid beamforming is based on using multiple antenna arrays for impartiality to get the multiplexing gain in mm-wave.

**5. Transmission Power and Power Consumption for Millimeter Wave Massive MIMO**

Research in [25] studied the high cost and power consumption to decrease the number of RF chains using multiplexing gain for large antenna arrays and hybrid precoding by using dual penalty decomposition. The simulation results in performed fully-digital precoding for an infinite resolution phase shifter. The author in [26] used novel hardware-efficient hybrid precoding to select the optimal antenna based on phase over-samplers and a switch to maximize hardware efficiency. The author in [26] achieved comparable spectral efficiency based on POS-SW-based hybrid beamforming. While [27] studied the novel massive MIMO system with transmitting array and reflect-array that contains little antenna of radio frequency chains to develop the sparsity of mm-wave channels. The proposed transmit array uses reflect-array to obtain low complexity of transmit power. The research [28] studied the analog and digital hybrid beamforming using the line-of-sight (LoS) channels, with Zero Forcing (ZF) at the baseband to achieve high spectral efficiency. The hybrid beamforming using the full digital ZF precoding provides minimal hardware complexity and lower power consumption.

TABLE 2. Summary for Minimizing Transmit Power Based on RF Chains

| Authors | Features | Limitation |
| --- | --- | --- |
| Qingjiang et al. | • Reducing the number of RF chains based spatial multiplexing gain of large-array. | • penalty dual decomposition (PDD)<br>• KKT solutions |
| Li et al. | • Reduced high hardware complexity and power consumption<br>• Performance spectral efficiency | • a limited number of simple phase over-samplers<br>• switch (SW) |
| Vahid et al. | • Developed a unified power consumption model based on a few active antennas | • reflect-array (RA)<br>• transmit - array (TA) antennas |
| Shehata et al. | • Used mathematically in pure line-of-sight (LoS) channels | • Zero Forcing (ZF) |

|  | • Using hybrid beamforming with Zero Forcing (ZF) at the baseband can achieve equivalent spectral efficiency |  |
|---|---|---|

## 6. Small Cells and Beamforming

The Dense heterogonous small cells are one of the key technologies which shall guarantee performance energy efficiency and improve high data rate in edge cell for 5G network by utilizing hybrid beamforming. Enhancing mobile traffic in massive MIMO for 5G Network depends on increasing the number of small cells and by connecting many small base stations with decreased cell sizes where the small cell provides a low-cost base station and a low power transmission. Small cells with massive MIMO can support a large number of distributed users by employing cell densification.

According to the the Friis transmission in [33], for the power gain $S_{ij} = s_{ij}^{tx} s_{ij}^{rx}$, $\lambda^2/16 \ \pi^2(\frac{d_{ij}}{d_0})$ where $s_{ij}^{tx} s_{ij}^{rx}$ are received and transmitted antennas, $d_0$ is the reference distance, $d_{ij}$ is the distance from $i$ to $j$ and η is the path losses exponents.

The signal-to-noise ratio transmitting form the base station to UE can be written as

$$\text{SINR}_{ij} = \frac{Q_{ij} S_{ij}}{\sum_{k=i, k \neq j} Q_{kj} S_{kj} + \sigma^2} \qquad (3)$$

where $Q_{ij}$ is the transmit power from the base station to UEs and $\sigma$ is the additive white Gaussian noise. The achievable data rates for UEs can be written as

$$R_{ij} = \frac{B}{K} \log_2(1 + SINR_{ij}) \qquad (4)$$

The small cell mitigates the inter-cell interference to achieve high spectral efficiency by using high directional beamforming. Reducing co-channel interference in dense small cells is chosen as a cell head to combine the intra-cell by using an unlicensed between macro-cell base station and dense small cells [29], [30]. The optimal beamforming for dens small cell (DSC) can be improved depending on real-time for the user scheduling and minimizing the total power consumption to achieve quality service

$$Q_{max} = \lim_{x_k, \forall k, j} Q_{dynamic} + Q_{static} \qquad (5)$$

$$\log_2(1 + \text{SINR}_k) \geq \delta_k \qquad \forall k \qquad (6)$$

$$\sum_j \varpi_{ij} = 1 \qquad \forall i \in Y \qquad (7)$$

where $\delta_k$ is a fixed quality of service, $Y$ is the distributed of UEs for every cell and $\varpi_{ij}$ is the connotation variable between UE $i$ and base station $j$. The total power constraint can be expressed as

$$\sum_i \varpi_{ij} Q_{ij} \leq Q_{max} \qquad (8)$$

In order to reduce the interference in small cell based on minimizing the transmit signal and the quality of service constraint can be written as

$$\sum_i \varpi_{ij} SINR_{ij} \geq R_f \tag{9}$$

where $R_f$ is the minimum achievable data rate for every UE. Reducing the propagation losses and mitigating the interference constraint can be expressed as:

$$\sum_i \varpi_{ij} Q_{ij} S_{ij} \leq \varphi_j \tag{10}$$

where $\varphi_j$ is the the highest interference constraint. Reducing the cost and energy consumption depend on selecting the optimal power allocation from the base station in cell j th to distributed UEs in cell i th for any subcarrier $\lambda$ according Eq.(2) as $Q_i^j(\lambda)$, based on reducing the path loss between the i th UEs and j th base station, the constraint power

$$Q_i \leq Q_i^{max} \tag{11}$$

To reduce the circuit power consumption, the total emitting power in term dynamic power was analysed which can be expressed as

$$Q_{dynamic} = \alpha \sum_{k=1}^{K} \|x_{k0}\|^2 + \sum_{j=1}^{r} \alpha_j \sum_{k=1}^{K} \|x_{kj}\|^2 \tag{12}$$

where $x_{k,0} \in^{M_{BS} \times 1}$ is the beamforming vector, and $x_{kj} \in^{M_{DSC} \times 1}$ is the beamforming to backhaul dense small cells $j = 0, \ldots, r$.

While the static power analysed to decrease the consumption circuit power by using several transmit antennas and model power dissipation for every antenna such as converting from analog to digital, or digital to analog converter, mixer, filters, and baseband processing. The power static can be written as

$$Q_{static} = \frac{\mathcal{T}_0}{\xi} M_{BS} + \sum_{j=1}^{r} \frac{\mathcal{T}_j}{\xi} M_{DSC} \tag{13}$$

where $\xi$ is the number of a subcarrier for $1 < j < r$, is the model power dissipation and $M_{DSC}$ is the number of transmit antennas in dense small cells. The lower propagation channel and high energy efficiency can be obtained by decreasing the distance in a dense small cell between active users and base station by reducing the high cost and energy consumption in order to obtain the minimum transmitted power as:

$$Q_{ij} = \frac{\varphi_{ij}}{S_{ij}}(2^{R_f} - 1) \tag{14}$$

## 7. Numerical Results

In this section, present numerical results were proposed to show the performance of total transmission power in dense small cells using Monte-Carlo simulations. Fig. 2 shows the average total transmit power consumption with the number of users, where the large hardware could be decreased by minimizing emitting transmit power and power dissipation from (12) and (13). The dense small cell provides a low-cost base station and a low power transmission based on distance from users to the transmitter. The dense small cell in massive MIMO can provide a high quality of service for a large number of distributed users based on employing cell densification. The high energy efficiency depends on total power consumption $Q_{dynamic} + Q_{static}$. Also, the low-cost hybrid beamforming can decrease hardware complexity, where every RF chain is connected with each antenna at a transmitting signal. From Fig. 2, the high transmit power per subcarrier proportional to the high quality of service. The high cost and large hardware complexity could be improved by avoiding devoting an unconnected RF chain for all antennas based on employing massive MIMO baseband precoding in dense small cells. By employing sufficient multiple antennas for the massive MIMO system in dense small cells with the distributed number of users, in this case, high energy efficiency can be achieved and the hardware complexity can decrease. The baseband beamforming in mm-wave is improved by decreasing propagation path loss among the information path and its estimation under constraints of the total transmission power $Q_{dynamic} + Q_{static}$.

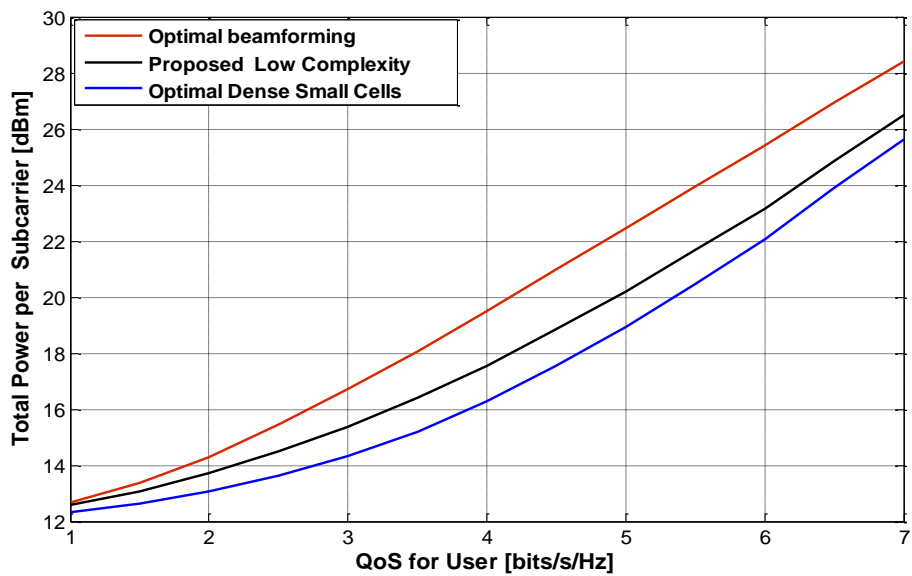

Figure 2. Related transmit power with a quality of service in dense small cells.

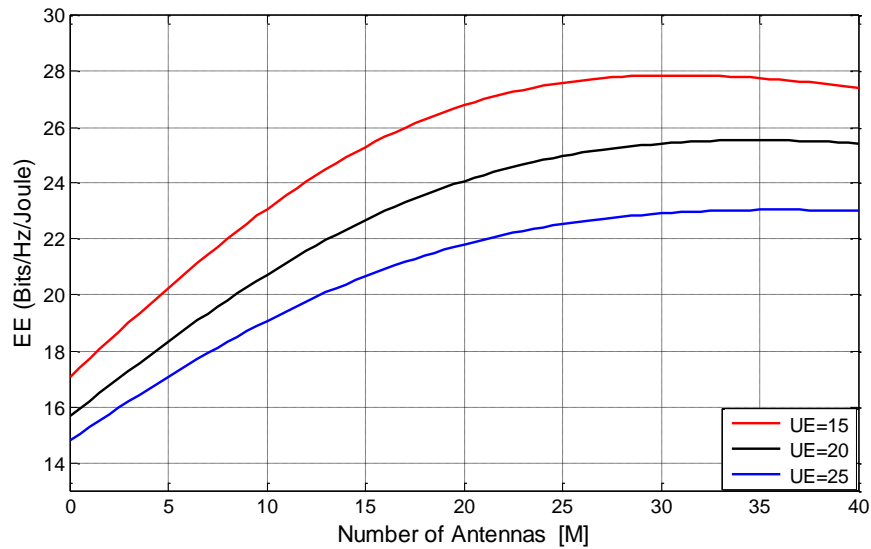

Figure 3. Energy efficiency versus transmit a number of antennas in small cells.

From Fig. 3, the maximum EE for number of antennas related to SINR constraint is showed. The EE is improved based on the load capacity and the minimum achievable data rate for every UE (9). In addition, depending on reducing the propagation losses and max-SINR can be expressed as (10). From Fig.3, the energy efficiency starts to increase when the maximum SINR is high by guaranteeing each UE connected with only one base station to avoid affecting the user association and reducing transmit power. While the energy efficiency starts decreasing when the number of antennas increases which require guarantee the low-cost and low-power based on providing a load-balancing property in order to increase the EE under the constraint power control and QoS requirements according to (9) and (11).

## 8. Conclusion

This paper presents massive MIMO with mm-wave beamforming to increase the quality of service by enhancing spectral and energy efficiencies. This article has explored several opportunities to improve energy efficiency and mitigated co-channel interference and improving a signal to interference noise ratio required. The hybrid beamforming can improve the channel, choose the minimum number of RF chains, reduce the hardware complexity and use the huge number of transmit antennas. This article also discussed low-cost hybrid digital-analog beamforming for multiuser massive MIMO by reducing the high cost and energy consumption based on using the minimum transmit power constraints to reduce hardware complexity for every RF chain. Reducing a large path loss in Rayleigh fading channel based on improving hybrid beamforming in mm-wave can exploit a large bandwidth. The discussion was showed clear indications that hybrid massive MIMO systems have a huge possibility to meet the energy efficiency demands expected in 5G cellular networks.

## 9. ACKNOWLEDGEMENTS

This research is supported by Ministry of Higher Education Malaysia through Fundamental Research Grant Scheme (FRGS/1/2019/TK04/UTHM/02/8) and partially sponsored by Universiti Tun Hussein Onn Malaysia under TIER1 Grant (vot H243).

## 10. References

[1] K. S. V. Prasad, E. Hossain, and V. K. Bhargava, "Energy efficiency in massive MIMO-based 5G networks: Opportunities and challenges," IEEE Wireless Communications. vol. 24, no. 3, pp. 86–94, 2017.

[2] R. Chen, H. Xu, C. Li, L. Zhu, and J. Li, "Hybrid beamforming for broadband millimeter wave massive MIMO systems," IEEE 87th Veh. Technol. Conf. (VTC Spring), pp. 1–5, 2018.

[3] N. Song, H. Sun, Q. Zhang, and T. Yang, "Channel alignment for hybrid beamforming in millimeter wave multi-user massive MIMO," 2017 IEEE Glob. Commun. Conf. GLOBECOM 2017 - Proc., vol. 2018-Jan., pp. 1–6, 2017.

[4] D. Castanheira, P. Lopes, A. Silva, and A. Gameiro, "Hybrid beamforming designs for massive MIMO millimeter-wave heterogeneous systems," IEEE Access, vol. 5, pp. 21806–21817, 2017.

[5] Z. Li, S. Han, and A. F. Molisch, "User-centric virtual sectorization for millimeter-wave massive MIMO downlink," 2017 IEEE Glob. Commun. Conf. GLOBECOM 2017 - Proc., vol. 2018-Jan., pp. 1–6, 2017.

[6] X. Zhang, A. F. Molisch, and S. Y. Kung, "Variable-phase-shift-based RF-baseband codesign for MIMO antenna selection," IEEE Trans. Signal Process., vol. 53, no. 11, pp. 4091–4103, 2005.

[7] P. Sudarshan, N. B. Mehta, A. F. Molisch, and J. Zhang, "Channel statistics-based RF pre-processing with antenna selection," IEEE Trans. Wirel. Commun., vol. 5, no. 12, pp. 3501–3510, 2006.

[8] S. Payami, M. Sellathurai, and K. Nikitopoulos, "Low-complexity hybrid beamforming for massive MIMO systems in frequency-selective channels," IEEE Access, vol. 7, no. c, pp. 36195–36206, 2019.

[9] M. Shehata, A. Mokh, M. Crussiere, M. Helard, and P. Pajusco, "On the equivalence between hybrid and full digital beamforming in mmwave communications," 26th Int. Conf. Telecommun. ICT 2019, April, pp. 391–395, 2019.

[10] F. Sohrabi and W. Yu, "Hybrid analog and digital beamforming for mmwave OFDM large-scale antenna arrays," IEEE J. Sel. Areas Commun., vol. 35, no. 7, pp. 1432–1443, 2017.

[11] X. Jiang and F. Kaltenberger, "Channel reciprocity calibration in TDD hybrid beamforming massive MIMO systems," IEEE J. Sel. Top. Signal Process., vol. 12, no. 3, pp. 422–431, 2018.

[12] X. Wu, D. Liu, and F. Yin, "Hybrid beamforming for multi-user massive MIMO systems," IEEE Trans. Commun., vol. 66, no. 9, pp. 3879–3891, 2018.

[13] A. Maltsev, R. Maslennikov, A. Sevastyanov, A. Khoryaev, and A. Lomayev, "Experimental investigations of 60 GHz WLAN systems in office environment," IEEE J. Sel. Areas Commun., vol. 27, no. 8, pp. 1488–1499, 2009.

[14] S. Geng, J. Kivinen, X. Zhao, and P. Vainikainen, "Millimeter-wave propagation channel characterization for short-range wireless communications," IEEE Trans. Veh. Technol., vol. 58, no. 1, pp. 3–13, 2009.

[15] N. T. T. Docomo, "5G channel model for bands up to100 GHz," Tech. Report, Oct. 2016.

[16] T. A. Thomas et al., "A prediction study of path loss models from 2-73.5 GHz in an urban-macro environment," IEEE Veh. Technol. Conf., vol. 2016-July, 2016.

[17] M. R. Akdeniz et al., "Millimeter wave channel modeling and cellular capacity evaluation," IEEE J. Sel. Areas Commun., vol. 32, no. 6, pp. 1164–1179, 2014.

[18] D. W. Yue, H. H. Nguyen, and S. Xu, "'Multiplexing analysis of millimeter-wave massive MIMO systems.,'" arXiv Prepr. arXiv1801.02987, 2018.

[19] W. Tan, X. Feng, G. Liu, W. Tan, M. Zhou, and C. Li, "Spectral efficiency of massive MIMO systems with multiple sub-arrays antenna," IEEE Access, vol. 6, no. c, pp. 31213–31223, 2018.

[20] H. Huang, Y. Song, J. Yang, G. Gui, and F. Adachi, "Deep-learning-based millimeter-wave massive MIMO for hybrid precoding," IEEE Trans. Veh. Technol., vol. 68, no. 3, pp. 3027–3032, 2019.

[21] D. Zhang, Z. Zhou, C. Xu, Y. Zhang, J. Rodriguez, and T. Sato, "Capacity analysis of NOMA with mmwave massive MIMO systems," IEEE J. Sel. Areas Commun., vol. 35, no. 7, pp. 1606–1618, 2017.

[22] Z. Li, S. Han, S. Sangodoyin, R. Wang, and A. F. Molisch, "Joint optimization of hybrid beamforming for multi-user massive MIMO downlink," IEEE Trans. Wirel. Commun., vol. 17, no. 6, pp. 3600–3614, 2018.

[23] W. Ni and X. Dong, "Hybrid block diagonalization for massive multiuser MIMO systems," IEEE Trans. Commun., vol. 64, no. 1, pp. 201–211, 2016.

[24] Ioushua, S. Stein, and Y. C. Eldar, "Hybrid analog-digital beamforming for massive mimo systems," rXiv Prepr. arXiv1712.03485, 2017.

[25] Q. Shi and M. Hong, "Spectral efficiency optimization for millimeter wave multiuser MIMO systems," IEEE J. Sel. Top. Signal Process., vol. 12, no. 3, pp. 455–468, 2018.

[26] M. Li, Z. Wang, H. Li, Q. Liu, and L. Zhou, "A hardware-efficient hybrid beamforming solution for mmwave MIMO systems," IEEE Wirel. Commun., vol. 26, no. 1, pp. 137–143, 2019.

[27] Jamali, V., T. A.M., Fischer, M. G., and R. and Schober, "Reflect-and transmit-array antennas for scalable and energy-efficient mmWave massive MIMO," arXiv Prepr. arXiv1902.07670, 2019.

[28] M. Shehata, A. Mokh, M. Crussiere, M. Helard, and P. Pajusco, "On the equivalence between hybrid and full digital beamforming in mmwave communications," 26th Int. Conf. Telecommun. ICT 2019, pp. 391–395, 2019.

[29] H. Q. Ngo, A. Ashikhmin, H. Yang, E. G. Larsson, and T. L. Marzetta, "Cell-free massive MIMO versus small cells," IEEE Trans. Wirel. Commun., vol. 16, no. 3, pp. 1834–1850, 2017.

[30] U. Habiba, H. Tabassum, and E. Hossain, Backhauling 5G small cells with massive- MIMO- enabled mmWave communication," Backhauling / Fronthauling Future. Wirel. Syst. Wiley, 2016.

[31] A. Salh, N. S. M. Shah, L. Audah, Q. Abdullah, W. A. Jabbar and M. Mohamad, "Energy-efficient power allocation and joint user association in multiuser-downlink massive MIMO system," in IEEE Access, vol. 8, pp. 1314-1326, 2020, doi: 10.1109/ACCESS.2019.2958640.

[32] Q. Abdullah, N. Abdullah, M. Balfaqih, N. S. M. Shah, S. Anuar, A. A. Almohammedi, and V. Shepelev, "Maximising system throughput in wireless powered sub-6 GHz and millimetre-wave 5G heterogeneous networks," Telkomnika (Telecommunication Computing Electronics and Control), vol. 18, no. 3, pp. 1185–1194, 2020.

[33] A. Salh, L. Audah, N. S. M. Shah, and S. A. Hamzah, "Maximizing energy efficiency in downlink massive MIMO systems by full-complexity reduced zero-forcing beamforming," International Journal of Engineering and Technology (UAE), vol. 7, no. 4, pp. 33–36, 2018.

[34] A. Salh, L. Audah, N. S. M. Shah, and S. A. Hamzah, "Trade-off energy and spectral efficiency in a downlink massive MIMO system," Wireless Personal Communications, vol. 106, no. 2, pp. 897–910, 2019.